\documentclass[12pt]{iopart}
\usepackage{graphicx,latexsym,amssymb}
\usepackage{dcolumn}
\usepackage{bm}
\usepackage{longtable}
\def\nA{nucleon-nucleus\ }

\def\nPb{$n+^{208}$Pb\ }
\def\pPb{$p+^{208}$Pb\ }
\def\pCa{$p+^{40}$Ca\ }

\begin{document}
\title{Rearrangement term in the nonlocal folding model of the nucleon optical 
 potential}
\author{Doan Thi Loan, Dao T. Khoa, and Nguyen Hoang Phuc}
\address{Institute for Nuclear Science and Technology, VINATOM \\ 
179 Hoang Quoc Viet, Cau Giay, 100000 Hanoi, Vietnam.}
\ead{khoa@vinatom.gov.vn} 
\begin{abstract}
{Based on the mean-field determination of the single-particle energy in 
nuclear matter that contains naturally a rearrangement term (RT) implied by the 
Hugenholtz-van Hove theorem, the folding model of the nucleon optical potential 
(OP) is extended to take into account the RT using the effective, density dependent 
CDM3Yn interaction. With the exchange part of the nucleon folded OP treated exactly 
in the Hartree-Fock manner, a compact nonlocal version of the folding model
is suggested in the present work to determine explicitly the isospin-dependent, 
nonlocal central term of the nucleon OP. To solve the optical model (OM) equation 
with a complex nonlocal OP, the calculable $R$-matrix method is used to analyse 
the elastic neutron and proton scattering on $^{40,48}$Ca, $^{90}$Zr, and $^{208}$Pb 
targets at low energies. The inclusion of the RT into the folding model calculation 
of the nonlocal nucleon OP was shown to be essential for the overall good OM description 
of elastic nucleon scattering. To validate the nonlocal version of the folding model, 
the OM results given by the nonlocal folded nucleon OP are also compared with those 
given by the global parametrization of the nonlocal OP using the analytical nonlocal 
form factor suggested by Perey and Buck.} 
\end{abstract}

\submitto{\jpg}
\maketitle

\section{Introduction}
\label{intro} 
Over a wide range of the single-particle (SP) energies, the nucleon motion in medium 
is overwhelmingly governed by the nuclear mean field, known as the shell-model 
potential for bound states and the optical potential (OP) for scattering states. 
The mean-field, SP potential is also the key quantity in the many-body studies 
of the equation of state of nuclear matter (NM) as well as the structure 
of finite nuclei \cite{Mah85,Bal07}. The nucleon OP in the NM limit has been well 
studied in the Brueckner-Hartree-Fock (BHF) calculations of NM using the free 
nucleon-nucleon (NN) interaction \cite{Bal07,Bom91,Zuo99,Zuo14}, or the mean-field 
calculation of NM on the Hartree-Fock (HF) level using different choices of the 
effective NN interaction \cite{Kho93,Kho95,Xu10,Che12,Loa15}. The mean-field 
prediction for the nucleon OP in the NM limit provides a vital input for the 
microscopic models of the nucleon OP of finite nuclei. In particular, the  
widely-used folding model of the nucleon OP (see, e.g., 
Refs.~\cite{Loa15,Kho02,Kho07,Minomo}).  

The microscopic many-body studies of NM have shown the important role by the
Pauli blocking effects, and the increasing strength of the three-body interaction
as well as other higher-order NN correlations at high densities of NM \cite{Bal07}. 
These in-medium effects are effectively taken into account by the density dependence 
explicit embedded in different versions of the effective NN interaction used in the 
nuclear structure and nuclear reactions studies. In the present work, we focus on 
the CDM3Yn density dependent versions \cite{Kho97} of the M3Y-Paris interaction 
\cite{An83} which have been successfully used in the HF studies of NM 
\cite{Kho93,Kho95,Kho96,Tha09,Loa11} as well as in the folding model calculation 
of the nucleon and nucleus-nucleus OP \cite{Kho02,Kho07,Kho97,Kho07r,Kho09,Kho14}. 
In general, the folding model calculation of the nucleon OP is done on the HF level, 
and the folded OP lacks, therefore, the higher-order \emph{rearrangement} term that 
arises naturally in the Landau theory of infinite Fermi systems \cite{Mig67}. 
Such a rearrangement term (RT) also presents in the SP potential when it is determined 
from the total energy of NM using the Hugenholtz-van Hove (HvH) theorem \cite{HvH,Sat99}, 
which is exact for all the interacting Fermi systems, independent of the interaction 
between fermions. In our recent HF study of NM \cite{Loa15}, the density dependence 
of the CDM3Yn interaction (with n=3,6) was modified to reproduce on the HF level 
the SP potential obtained from the total NM energy using the HvH theorem.  
A strong impact of the RT on the strength and shape of the folded nucleon 
OP was found \cite{Loa15} essential for a good optical model (OM) description of the 
elastic \nPb scattering at energies of 30.4 and 40 MeV. 

Because the standard local approximation \cite{Kho02} was used in Ref.~\cite{Loa15} 
to localize the (Fock-type) exchange term of the folded OP, it remains uncertain 
how the RT affects the OM results of elastic nucleon scattering when the antisymmetrization 
of the \nA system is exactly taken into account, and the folded nucleon OP becomes
{\em nonlocal}. Although some versions of the nonlocal folding model with the exact 
treatment of the exchange term are already available in the literature (see, e.g., 
Ref.~\cite{Minomo}), none of them has included the RT into the HF-type folding model
calculation. Therefore, the main goal of the present study is to explore the impact 
of the rearrangement contribution to the OM description of nucleon elastic scattering 
by the nonlocal folded nucleon OP that treats the exchange kernel exactly. For this 
purpose, a compact nonlocal version of the folding model of the nucleon OP is suggested, 
where the RT is taken into account using the modified density dependence of the CDM3Yn
interaction suggested in Ref.~\cite{Loa15}. As an important mean-field aspect of the SP 
potential, the RT is always explicitly taken into account in numerous variational HF 
calculations of nuclear structure using the effective density-dependent NN interaction. 
However, the RT has been so far neglected in most of the HF-type folding model 
calculations of the nucleon OP, i.e., the single-nucleon potential at positive energies. 
Our present research is expected to shed light on the important role of the RT in the 
OM description of elastic nucleon scattering using the nonlocal folded OP. Furthermore, 
a consistent folding model for the nonlocal nucleon mean-field potential including 
the RT should also be of interest for the studies of nuclear reactions at low energies, 
in particular, those relevant for the nuclear astrophysics, where the effect of the 
nonlocality of the \nA potential has been shown to be quite significant \cite{Bail17,Tian18}.   

In general, solving the Schr\"{o}dinger equation with a nonlocal potential readily
leads to an integro-differential equation that is more complicated to solve than 
a standard differential equation with a local potential. For the elastic nucleon 
scattering, the use of the nonlocal OP leads to an explicit angular-momentum 
dependence of the integral equation. At variance with the traditional methods 
for the solution of the integro-differential equation, we have chosen in the present 
work the calculable $R$-matrix method \cite{desco} to solve the OM equation 
with the nonlocal folded nucleon OP. This $R$-matrix method was recently extended 
\cite{desco, desco2} to include the Lagrange mesh and Gauss-Legendre quadrature 
integration that significantly simplify the numerical calculation. This method 
was tested in our recent OM analysis \cite{Loa18} of elastic nucleon scattering 
on different targets at energies up to 40 MeV, using the phenomenological nonlocal 
nucleon OP \cite{Perey,TPM,Lovel,Lovel2}. To validate the present nonlocal folding 
model of the nucleon OP, the OM results given by the nonlocal folded OP are also 
compared with those given by the global parametrization of the nonlocal nucleon OP 
suggested recently \cite{TPM,Lovel,Lovel2} using the analytical form of the 
Perey-Buck nonlocal form factor \cite{Perey}. 

\section{Single-particle potential in nuclear matter}
\label{sec1} 
\begin{figure}[bht]\vspace*{0cm}\hspace*{1cm}
\includegraphics[width=0.8\textwidth]{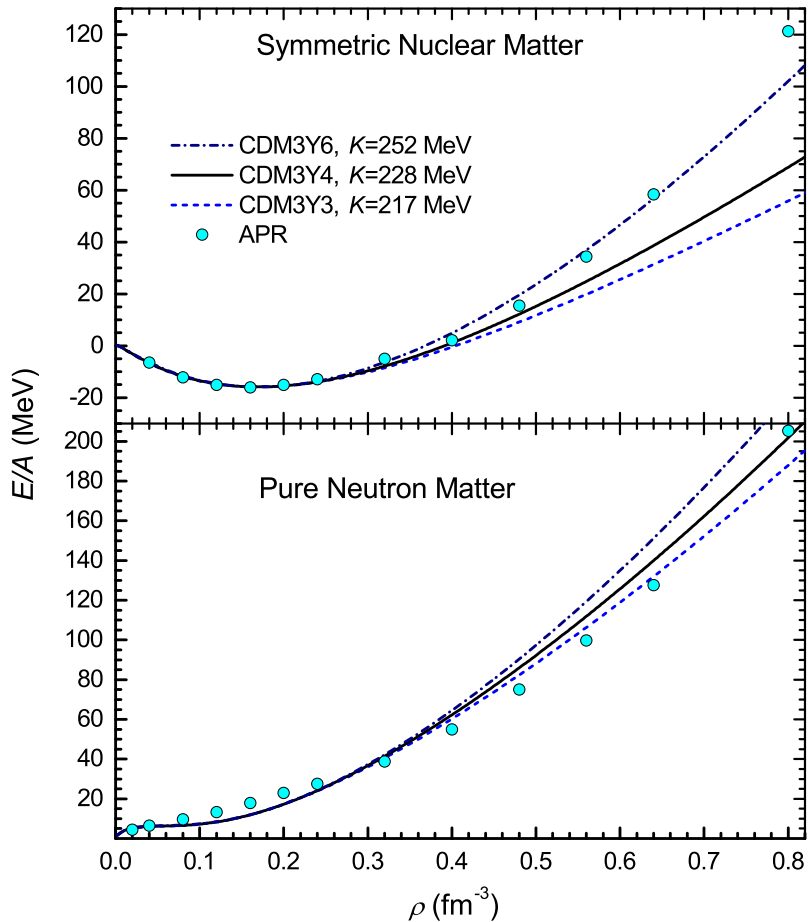}\vspace*{-1cm}
\caption{Energy per nucleon of the symmetric NM and pure neutron matter 
given by the HF calculation using the density dependent CDM3Yn interaction. 
$K$ is the nuclear incompressibility obtained at the saturation density
$\rho_0\approx 0.16$ fm$^{-3}$. The circles are results of the ab initio 
variational calculation by Akmal, Pandharipande and Ravenhall (APR) 
\cite{Ak98}.} \label{f1}
\end{figure}

Because an effective, density dependent NN interaction is the key input 
for the folding model calculation of the nucleon OP, we discuss first the 
density dependent CDM3Yn interaction which is based on the mean-field 
description of the SP potential in NM. Originally, parameters of the 
density dependence of the CDM3Yn interaction were parametrized \cite{Kho97} 
to reproduce the saturation properties of symmetric NM in the HF calculation. 
Later on, these parameters were updated to include a realistic isovector part 
\cite{Loa15,Kho07}. On the HF level, the CDM3Yn interaction is proven to give 
a good description of the EOS of NM \cite{Loa11}. As an illustration, we show 
in Fig.~\ref{f1} the HF results for the NM energy per nucleon obtained with the 
CDM3Yn interaction in comparison with the results of the \emph{ab initio} 
variational calculation of NM using the Argonne V18 interaction \cite{Ak98}. 
One can see a nice agreement of the HF results with those of the ab initio calculation,
especially, at low densities up to the saturation density $\rho_0\approx 0.16$ fm$^{-3}$
which is known to be accessible by elastic \nA scattering. In the present work, we 
focus particularly on the impact of the RT in the folding model description of the 
nonlocal nucleon OP, given a significant contribution by the RT to the \emph{local}
folded nucleon OP shown in Ref.~\cite{Loa15}.  

In general, according to Landau theory for an infinite system of interacting 
fermions \cite{Mig67}, the single-nucleon energy is determined \cite{Loa15} as 
the derivative of the energy per nucleon $\varepsilon\equiv E/A$ of NM  
with respect to the nucleon momentum distribution $n_\tau(k)$ as  
\begin{equation}
 E_{\tau}(\rho,k)=\frac{\partial \varepsilon}{\partial n_\tau(k)}=
\frac{\hbar^2 k^2}{2m_\tau}+U_{\tau}(\rho,k),\ {\rm where}\ \tau=n\ 
{\rm or}\ p. \label{eq1}
\end{equation}
$E_{\tau}(\rho,k)$ is, thus, the change of the energy of NM at the nucleon density 
$\rho$ caused by the removal or addition of a nucleon with the momentum $k$. 
The SP potential $U_{\tau}(\rho,k)$ consists of both the HF and rearrangement 
terms 
\begin{equation}
 U_{\tau}(\rho,k)= U^{\rm (HF)}_{\tau}(\rho,k)+U^{\rm (RT)}_{\tau}(\rho,k). 
 \label{eq2}
\end{equation}
The explicit expressions of $U_{\tau}(k)$ obtained in the HF calculation of NM
using a density dependent NN interaction have been given in Ref.\cite{Loa15}. 
At the Fermi momentum $(k\to k_F),\ E_\tau(k_F)$ determined from Eqs.~(\ref{eq1}) 
and (\ref{eq2}) is exactly the Fermi energy given by the Hugenholtz-van Hove (HvH)
theorem \cite{HvH}. 
We note that the HvH theorem is satisfied on the HF level only when the effective  
NN interaction is density independent, with the RT equal zero \cite{Loa15,Cze02}. 
In the mean-field calculation (\ref{eq1})-(\ref{eq2}) of the SP potential, the RT 
originates naturally from the density dependence of the effective NN interaction that 
presumably accounts for the higher-order NN correlations as well as the three-body 
force. Indeed, the RT was shown in the BHF studies of the SP potential 
in NM \cite{Mah85,Zuo99,Zuo14} to be due to the higher-order terms like the 
second-order diagram in the perturbative expansion of the mass operator or 
contribution from the three-body forces. 

For the spin-saturated NM, the direct (D) and exchange (EX) terms of the central 
part of the CDM3Yn interaction \cite{Kho97,Kho07} are used explicitly in the HF 
calculation of the SP potential in NM 
\begin{equation}
 v_{\rm D(EX)}(\rho,s)=F_0(\rho)v^{\rm D(EX)}_{00}(s) + 
 F_1(\rho)v^{\rm D(EX)}_{01}(s) \bm{\tau}_1\cdot\bm{\tau}_2. \label{eq3}
\end{equation}  
The radial parts of the isoscalar (IS) and isovector (IV) two-body force 
$v^{\rm D(EX)}_{00(01)}(s)$ are kept unchanged as determined from the original 
M3Y-Paris interaction \cite{An83}, in terms of three Yukawas. The parameters of the IS 
density dependence $F_0(\rho)$ were determined in the HF calculation \cite{Kho97} 
to reproduce the empirical saturation point of symmetric NM, with the nuclear 
incompressibility $K$ around 230 MeV (see Fig.~\ref{f1}). The parameters of the IV 
density dependence $F_1(\rho)$ were determined and fine tuned \cite{Loa15} 
by the isospin dependence of nucleon OP in asymmetric NM given by the BHF 
calculation by Jeukenne, Lejeune and Mahaux (JLM) \cite{Je77} and folding
model description of the charge exchange $(p,n)$ scattering to the isobar analog 
states in medium-mass nuclei \cite{Kho14}. Based on the exact expression of the RT 
of the SP potential given by the HvH theorem at different densities of NM, a compact 
method was suggested \cite{Loa15} to account for the RT on the HF level. Namely, 
the density dependence of the CDM3Yn interaction is added by a correction term 
originated from the RT, $F_{0(1)}(\rho)\to F_{0(1)}(\rho)+\Delta F_{0(1)}(\rho)$, 
so that the total SP potential can be calculated on the standard HF level 
\begin{equation}
\hskip -1.5cm U_{\tau}(\rho,k)=\sum_{k'\sigma'\tau'} 
 \langle \bm{k}\sigma\tau,\bm{k}'\sigma'\tau'|v_{\rm D}|
 \bm{k}\sigma\tau,\bm{k}'\sigma'\tau'\rangle +  
 \langle \bm{k}\sigma\tau,\bm{k}'\sigma'\tau'|v_{\rm EX}|
 \bm{k}'\sigma\tau,\bm{k}\sigma'\tau'\rangle,  \label{eq4}
\end{equation}
where $|\bm{k}\sigma\tau\rangle$ are the ordinary plane waves. Treating explicitly
the isospin dependence, the SP potential (\ref{eq4}) can be expressed \cite{Loa15} 
in terms of the IS and IV components as 
\begin{equation}
\hskip -1.5cm U_\tau(\rho,k)=[F_0(\rho)+\Delta F_0(\rho)]U^{\rm (M3Y)}_0(\rho,k)
\pm [F_1(\rho)\pm \Delta F_1(\rho)]U^{\rm (M3Y)}_1(\rho,k), \label{eq5}
\end{equation}
where (-) sign pertains to $\tau=p$ and (+) sign to $\tau=n$. $U^{\rm (M3Y)}_{0}$ and 
$U^{\rm (M3Y)}_1$ are the IS and IV parts of the SP potential, respectively, given 
by the HF calculation of NM using the original density independent M3Y interaction. 
More details on the density dependent functions $F_{0(1)}(\rho), \Delta F_{0(1)}(\rho)$,
and $U^{\rm (M3Y)}_{0(1)}(\rho,k)$ are given in Ref.~\cite{Loa15}. Because the original 
M3Y interaction is momentum independent, the momentum- or energy dependence of the SP 
potential (\ref{eq5}) is entirely determined by the exchange terms of $U^{\rm (M3Y)}_{0(1)}$. 
It is obvious from Eq.~(\ref{eq1}) that the in-medium nucleon momentum $k$ is 
determined self-consistently from the SP energy $E_\tau$ as 
\begin{equation}
 k=\sqrt{\frac{2m_\tau}{\hbar^2}[E_\tau(\rho,k)-U_\tau(\rho,k)]}. \label{eq6}
\end{equation}

\begin{figure}[bht] \vspace*{0cm}\hspace*{1cm}
\includegraphics[width=0.9\textwidth]{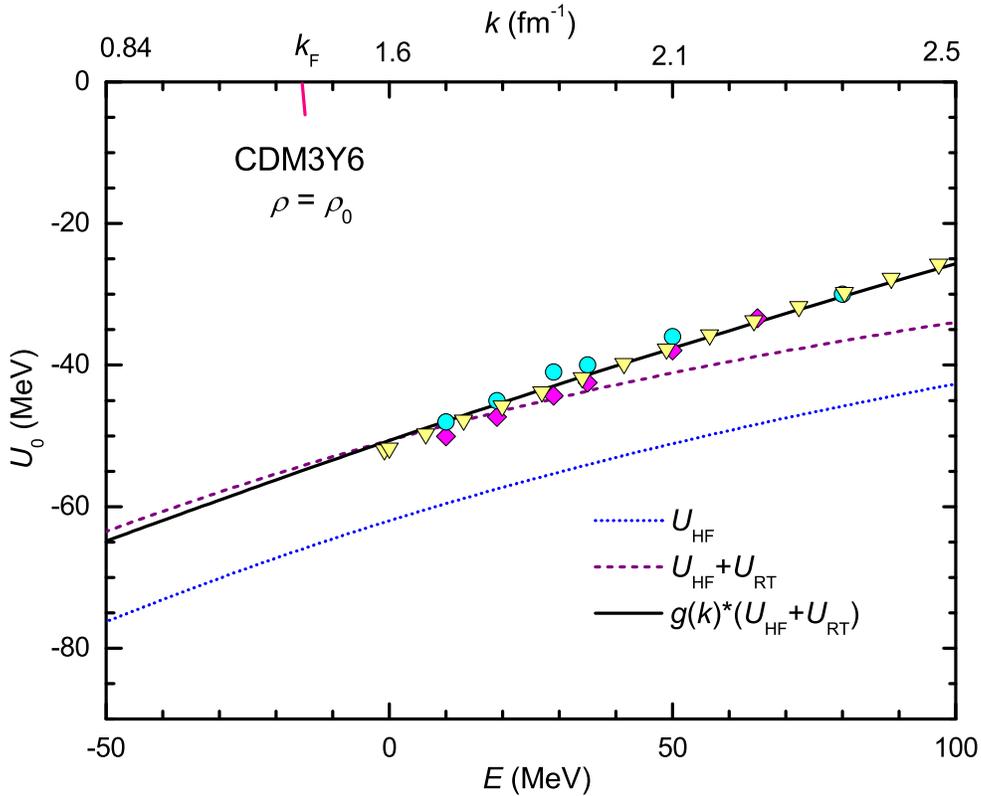}\vspace*{0cm}
 \caption{Single-nucleon potential in symmetric NM  (\ref{eq7}) determined 
at $\rho\approx \rho_0$ with and without the RT using the CDM3Y6 interaction, in 
comparison with the empirical data for the nucleon OP taken from 
Refs.~\cite{BM69} (circles), \cite{Var91} (squares), and \cite{Hama} (triangles). 
The momentum dependent factor $g(k)$ was obtained \cite{Loa15} by the $\chi^2$ 
fit of the calculated potential (\ref{eq7}) at $E>0$ to the empirical data 
(solid line).} \label{f2}
\end{figure}

With the density dependence of the CDM3Yn interaction fine tuned to reproduce the 
saturation properties of NM as shown in Fig.~\ref{f1}, the present HF approach 
provides a continuous description of both the SP potential for nucleons bound in NM, 
$U_{\tau}(\rho,k)$ with $k<k_F$, and the nucleon optical potential, $U_{\tau}(\rho,k)$ 
with $k>k_F$. This is the well-known \emph{continuous approximation} for the 
SP potential \cite{Mah85,Mah91}, where the nucleon OP in NM is determined as 
the mean-field potential felt by a nucleon incident on NM at the energy $E>0$. 
Complying with the Landau theory for a system of interacting fermions \cite{Mig67}, 
the nucleon OP (or the SP potential at positive energies) is also determined by the  
relations (\ref{eq1}) and (\ref{eq2}), so that the nucleon OP consists again 
of both the HF term and RT. The IS part of the nucleon OP, i.e., the nucleon OP 
in symmetric NM is determined as  
\begin{eqnarray}
\hskip -1cm U_0(E,\rho)&=& [F_0(\rho)+\Delta F_0(\rho)]
 U^{\rm (M3Y)}_0\big(\rho,k(E,\rho)\big) \nonumber\\
 &=& [F_0(\rho)+\Delta F_0(\rho)]\left[J^D_0+\int\hat j_1(k_Fr)j_0
 \big(k(E,\rho)r\big)v^{\rm EX}_{00}(r)d^3r\right], \label{eq7} 
\end{eqnarray}
\begin{equation}
\hskip -1cm {\rm where}\ J^{\rm D}_0=\int v_{00}^{\rm D}(r)d^3r, \
\hat{j_1}(x)=3j_1(x)/x\ =\ 3(\sin x-x\cos x)/x^3. \nonumber 
\end{equation}
$k(E,\rho)$ is the in-medium momentum of the incident nucleon propagating 
in the mean field of bound nucleons in NM, and is determined by the same 
relation (\ref{eq6}) but with $E_\tau$ replaced by the incident energy $E$. 
Within the time-independent HF formalism, the energy- and momentum dependences 
of the nucleon OP are treated on the same footing via the relation (\ref{eq6}) 
as illustrated in Fig.~\ref{f2}. 
Therefore, an important constraint for the present study is that at $E>0$ the energy 
dependence of the potential (\ref{eq7}) should agree reasonably with the observed 
energy dependence of the nucleon OP. The total SP potential in symmetric NM (\ref{eq7}) 
evaluated at $\rho_0$ using the CDM3Y6 interaction is compared with the empirical 
data \cite{BM69,Var91,Hama} in Fig.~\ref{f2}. One can see that the inclusion of the RT 
significantly improves the agreement with the empirical data at low energies ($E<50$ MeV). 
The HF results shown in Fig.~\ref{f1} also confirms that at low energies the energy
dependence of the nucleon OP is mainly determined by the Fock-type exchange term
of the nucleon OP, i.e., by the antisymmetrization effect. At higher energies, 
the agreement worsens, and this is a well expected effect, because the energy 
dependence of the nucleon OP in NM was shown in the microscopic BHF calculation 
\cite{Mah91} to originate not only from the exchange part, but also from the direct 
part of the Brueckner G-matrix. To have a realistic momentum dependence of the nucleon 
OP at higher energies or momenta, the nucleon OP given by the HF calculation was 
scaled \cite{Loa15} by a momentum dependent function $g(k)$ determined from the $\chi^2$ 
fit to the empirical data, $U(\rho,E)=g(k(E,\rho))U_0(\rho,E)$ (see Fig.~\ref{f2}). 
In the present work, we focus on the folding model analysis of elastic nucleon 
scattering at low energies ($E\leqslant 45$ MeV), and assume $g(k)\approx 1$ in the 
folding calculation of the nucleon OP.

\section{Folding model of the nucleon optical potential}
\subsection{Nucleon folded OP with a nonlocal exchange kernel}
\label{sec2} 
The folding model of the nucleon OP is known to generate the first-order 
term of the microscopic nucleon OP within the Feshbach's formalism 
of nuclear reactions \cite{Fe92}. The success of the folding approach in the 
description of elastic \nA scattering at low and medium energies confirms that the 
first-order term of the Feshbach's microscopic OP is indeed the dominant part 
of the nucleon OP. Applying the \emph{local density approximation} (LDA), commonly 
adopted in the HF calculations of nuclear structure, the plane waves 
$|\bm{k}'\sigma'\tau'\rangle$ in the SP potential (\ref{eq4}) can be replaced
by the SP wave functions $|j\rangle$ of the target nucleons. Then, the central 
OP of the elastic nucleon scattering on the target $A$ can be evaluated as
\begin{equation} 
  U(k)=\sum_{j\in A}[\langle \bm{k},j|v_{\rm D}|\bm{k},j\rangle 
	+\langle \bm{k},j|v_{\rm EX}|j,\bm{k}\rangle]. \label{eq8}
\end{equation}
The antisymmetrization of the \nA system is done in the HF manner, taking into 
account explicitly the knock-on exchange. As a result, the exchange term of the 
\nA potential (\ref{eq8}) becomes \emph{nonlocal}, and the OM equation for the 
elastic nucleon scattering at the energy $E$ is an integro-differential equation 
\begin{eqnarray}
 \left[-\frac{\hbar^2}{2\mu}\nabla^2+U_{\rm D}(R)+V_{\rm C}(R)
 +V_{\rm s.o.}(R)(\bm{l}.{\bm{\sigma}})\right]\chi(\bm{R}) 
 \nonumber\\ \hskip 2.6cm +\int K(\rho,\bm{R},\bm{r})
 \chi(\bm{r})d^3r= E\chi(\bm{R}), \label{eq9}
\end{eqnarray}
where $V_{\rm s.o.}(R)$ is the spin-orbit potential, and the Coulomb potential 
$V_{\rm C}(R)$ is included only for elastic proton scattering. The scattering 
wave function $\chi(\bm{R})$ is obtained from the solution of the OM equation 
(\ref{eq9}) at each \nA distance $R$. The energy dependent mean-field part 
consists of the local direct potential $U_{\rm D}(R)$ and the exchange integral 
with a nonlocal, density dependent kernel $K(\rho,\bm{R},\bm{r})$. 
The mean-field part of the nucleon OP can be expressed, in a manner consistent 
with the Lane representation, in terms of the isoscalar and isovector parts as   
\begin{eqnarray}
 && U_{\rm D}(R)=U^{\rm D}_{\rm IS}(R)\pm U^{\rm D}_{\rm IV}(R), \nonumber\\ 
 && K(\rho,\bm{R},\bm{r})=K_{\rm IS}(\rho,\bm{R},\bm{r})
 \pm K_{\rm IV}(\rho,\bm{R},\bm{r}), \label{eq10}  
\end{eqnarray}
where (-) sign pertains to proton OP and (+) sign to neutron OP. The IS and IV 
terms in Eq.~(\ref{eq10}) are determined using, respectively, the IS and IV parts 
of the nucleon density matrices as  
\begin{eqnarray}
&& U^{\rm D}_{\rm IS(IV)}(R)=
\int\big[\rho_n(\bm{r})\pm\rho_p(\bm{r})\big]v^{\rm D}_{00(01)}(\rho,s)d^3r, 
 \nonumber \\  
&& K_{\rm IS(IV)}(\rho,\bm{R},\bm{r})=
\big[\rho_n(\bm{R},\bm{r})\pm\rho_p(\bm{R},\bm{r})\big]
v^{\rm EX}_{00(01)}(\rho,s), \label{eq11}
\end{eqnarray}
where $s=|\bm R-\bm r|$. The nucleon density matrix is determined from the SP 
wave functions of target nucleons as
\begin{equation}
 \rho_\tau(\bm{r},\bm{r}')=\sum_{j\in A}\varphi^{(\tau)*}_j(\bm{r})
 \varphi^{(\tau)}_j(\bm{r}'),\ {\rm with}\ 
 \rho_\tau(\bm r)\equiv \rho_\tau(\bm{r},\bm{r}),\ {\rm and}  
 \ \tau=n,p. \label{eq12}
\end{equation}
Within the adopted LDA, the parameters of the density dependence of the CDM3Yn 
interaction determined in the HF calculation of NM are readily used in the HF-type 
folding model calculation of the nucleon OP of finite nuclei (\ref{eq8}), where 
the density dependent functional $F_{0(1)}(\rho)+\Delta F_{0(1)}(\rho)$ is given 
consistently by the local target density $\rho(\bm r)$ appearing in Eqs.~(\ref{eq10}) 
and (\ref{eq11}). The direct potential $U_{\rm D}(R)$ is obtained simply by folding the 
local nucleon density matrices with the direct part $v^{\rm D}_{00(01)}(\rho,s)$ 
of the density dependent CDM3Yn interaction (see more details in Ref.~\cite{Kho02}), 
including the contribution of the rearrangement term. We show here the explicit 
expression of the IV part of the direct folded potential 
\begin{equation} \label{eq13}
U^{\rm D}_{\rm IV}(R)=\int\big[\rho_n(\bm{r})-\rho_p(\bm{r})\big]
\big[F_1(\rho(\bm r))\pm\Delta F_1(\rho(\bm r))\big]v^{\rm D}_{01}(s)d^3r,
\end{equation}
where the $(\pm)$ signs are used in the same way as in Eq.~(\ref{eq10}). One 
can see that the contribution of the RT to the IV part of the direct potential 
$U^{\rm D}_{\rm IV}$ via $\Delta  F_1(\rho)$ is the same for both the proton and 
neutron OP. 

Note that the energy dependence of the folded nucleon OP is implicitly embedded 
in the exchange integral of Eq.~(\ref{eq9}) when the nonlocal exchange term is 
treated exactly. This procedure is cumbersome and involves the explicit angular-momentum 
dependence of the exchange kernel. Using the multipole decomposition of the radial Yukawa 
function in the exchange component of the CDM3Yn interaction (\ref{eq3}) 
\begin{equation}  
 v^{\rm EX}_{00(01)}(s)=\sum_{\lambda\mu}\frac{4\pi}{2\lambda+1}
 X^{(\lambda)}_{00(01)}(R,r)Y^*_{\lambda\mu}(\hat{\bm R})
 Y_{\lambda\mu}(\hat{\bm r}), \label{eq14} 
\end{equation}
we obtain, after integrating out the angular dependence, the following radial OM 
equation for each partial wave $L$
\begin{eqnarray} 
&&-\frac{\hbar^2}{2\mu}\left[\frac{d^2}{dR^2}-\frac{L(L+1)}{R^2}\right]
\chi_{LJ}(R)+\big[U_{\rm D}(R)+V_{\rm C}(R) \nonumber\\
&&+A_{LJ}V_{\rm s.o.}(R)\big]\chi_{LJ}(R) +\int_0^\infty K_{LJ}(\rho,R,r)
\chi_{LJ}(r)dr=E\chi_{LJ}(R), \label{eq15}
\end{eqnarray}
where the s.o. coupling coeficient $A_{LJ}=L$ if $J=L+1/2$, and 
$A_{LJ}=-L-1$ if $J=L-1/2$. One can see that the use of the nonlocal OP leads to 
an explicit angular-momentum dependence of the integral equation for the scattering 
wave function. The nonlocal exchange kernel is determined explicitly as  
\begin{eqnarray}
 \hskip -2cm K_{LJ}(\rho,R,r)=\big[K^{\rm IS}_{LJ}(\rho,R,r)
 \pm K^{\rm IV}_{LJ}(\rho,R,r)\big], \label{eq16} \\
 \hskip -2cm K^{\rm IS}_{LJ}(\rho,R,r)=\big[F_0(\rho(r))+
 \Delta F_0(\rho(r))\big]\sum_{nlj,\lambda}\frac{u^{(n)}_{nlj}(R)u^{(n)}_{nlj}(r)
 +u^{(p)}_{nlj}(R)u^{(p)}_{nlj}(r)}{R~r} \nonumber\\ 
 \hskip 1.5cm \times(2j+1) X^{(\lambda)}_{00}(R,r)
 \left(\begin{array}{l}L\\0\end{array}
 \begin{array}{l}l\\0\end{array}
 \begin{array}{l}\lambda\\0\end{array}\right)^2, \label{eq17} \\ 
 \hskip -2cm K^{\rm IV}_{LJ}(\rho,R,r)=\big[F_1(\rho(r))\pm
 \Delta F_1(\rho(r))\big]\sum_{nlj,\lambda}\frac{u^{(n)}_{nlj}(R)u^{(n)}_{nlj}(r)
 -u^{(p)}_{nlj}(R)u^{(p)}_{nlj}(r)}{R~r} \nonumber\\ 
 \hskip 1.5cm \times(2j+1) X^{(\lambda)}_{01}(R,r)
 \left(\begin{array}{l}L\\0\end{array}
 \begin{array}{l}l\\0\end{array}
 \begin{array}{l}\lambda\\0\end{array}\right)^2. \label{eq18}  
\end{eqnarray}
Here $u^{(\tau)}_{nlj}(r)$ is the radial part of the SP wave function
$\varphi^{(\tau)}_{nlj}(\bm{r})$ of the target nucleon. Note that in 
Eqs.~(\ref{eq16}) and (\ref{eq18}), the (-) sign is used with the proton OP 
and (+) sign with the neutron OP. Thus, the contribution of the RT to the IV 
part of the exchange kernel is also the same for both the proton and neutron OP 
as found for the IV part of the direct potential (\ref{eq13}). The explicit 
representation of the nucleon OP in terms of the IS and IV parts is helpful 
for the investigation of the contribution of valence neutrons to the OP. 
Furthemore, the form factor (FF) of the charge exchange $(p,n)$ reaction to the 
isobar analog state (IAS) is determined, in the Lane isospin coupling scheme, 
entirely by the IV part of the nucleon OP \cite{Kho14,Loc14}. Therefore, 
the present nonlocal folding model can be used to calculate the nonlocal 
charge exchange FF in the future folding model studies of the $(p,n)$ reaction 
to IAS.     

\subsection{Local approximation for the folded nucleon OP}
\label{sec3}
Although it is established that the nucleon OP is nonlocal in the coordinate 
space due to the Pauli blocking (as shown above) and the multichannel coupling, 
over the years the nucleon OP is mainly assumed in the local form for the OM 
analysis of elastic nucleon scattering. The local OP that describes properly the 
nucleon elastic scattering is the key input for the distorted wave Born approximation 
(DWBA) or coupled channel (CC) analyses of different direct reaction processes induced 
by the incident nucleon. Here the (bare) real OP accounts for the purely elastic 
scattering and imaginary OP accounts for the absorption of flux by those nonelastic 
reaction channels that are not explicitly taken into account in the CC calculation.
It is of interest, therefore, to assess the accuracy of the local version 
\cite{Kho02,Kho07} of the nonlocal folding model suggested in the present work.

Applying a local WKB approximation \cite{Sin75,Bri77} for the change in the 
scattering wave function in the OM equation (\ref{eq9}) induced by the exchange 
of spatial coordinates of the incident nucleon and that bound in the target 
\begin{equation}
\chi(\bm{r})=\chi(\bm{R+s})\simeq \chi(\bm{R})
 \exp\big(i{\bm k}(E,\bm R).{\bm s}\big), \label{eq19} 
\end{equation}
the exchange integral in Eq.~(\ref{eq9}) can be evaluated independently using 
the nonlocal nucleon density matrix. This gives rise to a \emph{local} 
exchange term of the folded \nA potential (\ref{eq8}) that depends explicitly
on energy via the local momentum of the incident nucleon $k(E,R)$  
\begin{eqnarray}
 \hskip -1cm U_{\rm EX}(E,R)=U^{\rm EX}_{\rm IS}(E,R)\pm 
  U^{\rm EX}_{\rm IV}(E,R), \nonumber\\ 
 \hskip -1cm U^{\rm EX}_{\rm IS(IV)}(E,R)=
  \int\big[\rho_n(\bm{R},\bm{r})\pm\rho_p(\bm{R},\bm{r})\big] 
  j_0\big(k(E,R)s\big) v^{\rm EX}_{00(01)}(\rho,s)d^3r, \label{eq20}
\end{eqnarray}
where the $(\pm)$ signs are used in the same way as in Eq.~(\ref{eq10}). The local 
momentum $k(E,R)$ of the incident nucleon propagating in the target mean field is 
determined from the real part of the total folded potential 
$U(E,R)=U_{\rm D}(R)+U_{\rm EX}(E,R)$ as
\begin{equation}
 k^2(E,R)=\frac{2\mu}{\hbar^2}[E-{\rm Re}~U(E,R)-V_{\rm C}(R)].\label{eq21}
\end{equation}  
The method used to evaluate the direct (\ref{eq11}) and exchange (\ref{eq20}) 
folded potentials has been discussed earlier (see, e.g., Ref.~\cite{Kho02}). 
Using a realistic local approximation for the nonlocal density matrix in the 
exchange potential (\ref{eq20}), the nuclear density $\rho(\bm r)$ obtained in 
any structure model or directly deduced from the electron scattering data 
can be used in the folding calculation of the nucleon OP. 
A preliminary folding model study of elastic \nPb scattering using the neutron 
OP obtained with the local approximation (\ref{eq20}) for the exchange term 
has shown a significant contribution of the RT to the nucleon folded OP 
\cite{Loa15}. It should be noted that the RT is commonly taken into account 
in the variational HF calculation of nuclear structure using some density 
dependent NN interaction. However, the RT has not been included so far in most 
of the HF-type folding model calculations of the nucleon OP, and the main goal 
of the present work is to show the impact of the RT on both the local and  
nonlocal folded nucleon OP.

\section{Elastic nucleon scattering on $^{40,48}$Ca, $^{90}$Zr,
 and $^{208}$Pb targets}
\label{sec4} 
The new version of the folding model for the nonlocal and local nucleon OP
discussed above in sections \ref{sec2} and \ref{sec3} has been used in the present
work to calculate the nucleon OP for the OM study of the elastic neutron and 
proton scattering on $^{40,48}$Ca, $^{90}$Zr, and $^{208}$Pb targets. 
One can see from Eqs.~(\ref{eq11})-(\ref{eq12}) that the folding calculation of the
\emph{nonlocal} OP requires explicitly the single-particle wave functions of all 
nucleons bound in the target. We have used here the SP wave functions given by the 
HF calculation of finite nuclei, using a complete basis of spherical Bessel functions 
\cite{Tha11} and the finite-range D1S Gogny interaction \cite{Ber91}.     

In the context of a \emph{complex} folded OP, it is necessary to have a realistic 
complex parametrization of the density dependent CDM3Yn interaction. 
For this purpose, the imaginary density dependence of the CDM3Yn interaction 
was determined using the same density dependent functionals $F_{0(1)}(\rho)$ as those 
used in Eq~(\ref{eq3}) for the real interaction, with the parameters determined at each 
energy to reproduce on the HF level \cite{Kho07} the energy dependent imaginary 
nucleon OP given by the JLM parametrization of the BHF results for NM \cite{Je77}. 
The folded complex nonlocal nucleon OP as well as its local version were further used 
as the input for the OM calculation of elastic nucleon scattering using the extended 
$R$-matrix method \cite{desco}. In the present OM calculation, the nonlocal mean-field 
part of the nucleon OP is supplemented by the local Coulomb and spin-orbit potentials 
taken from the global systematics CH89 of the nucleon OP \cite{Var91}. 

The reliability of the folded OP in the OM study of elastic nucleon scattering is 
best to be probed in the analysis of elastic neutron scattering from a heavy target 
at low energies, where the Coulomb interaction is absent and the mean-field dynamics 
is well established. The elastic \nPb scattering data accurately measured over a wide 
angular range at energies of 26, 30.4, and 40 MeV \cite{nPb1,nPb2,nPb3} turned out 
to be a very good test ground for this purpose. Because the parameters of the (real) 
CDM3Yn interaction were 
\begin{figure}[bht] \vspace*{-0.5cm}\hspace*{0cm}
\includegraphics[width=1.1\textwidth]{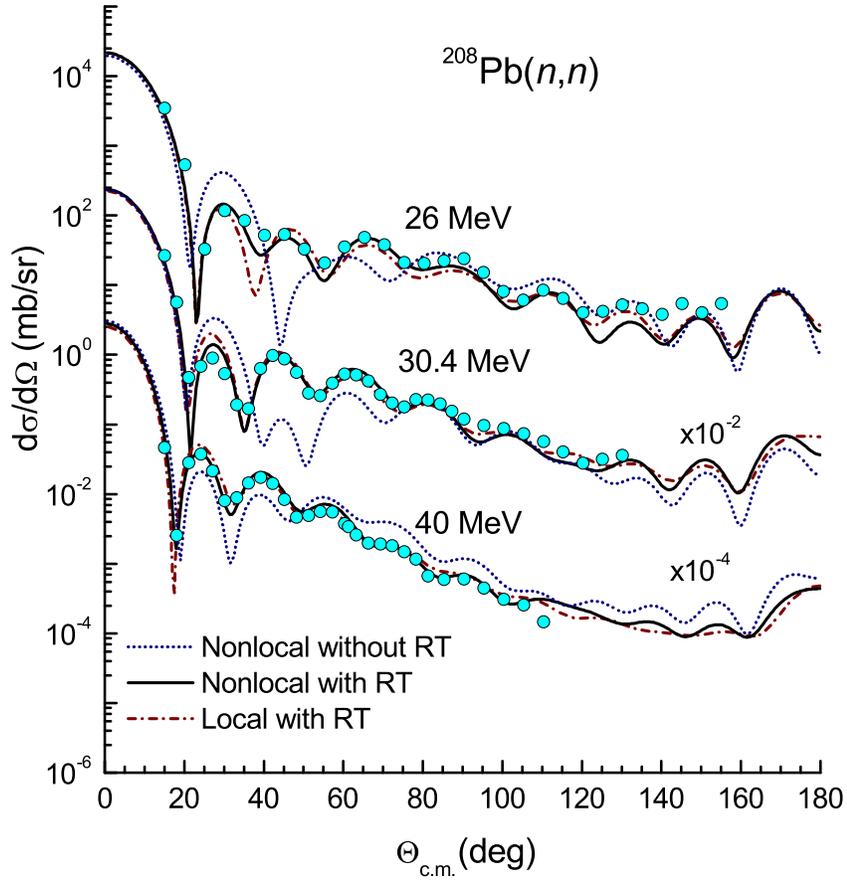}\vspace*{-3.5cm}
 \caption{OM description of the elastic \nPb scattering data measured at 26, 30, 
and 40 MeV \cite{nPb1,nPb2,nPb3} given by the complex \emph{nonlocal} folded OP 
obtained with the CDM3Y6 interaction with or without the inclusion of the RT, 
in comparison with that given by the \emph{local} folded OP with the RT 
included.} \label{f3}
\end{figure}
adjusted by the realistic HF description of NM as shown in Fig.~\ref{f1}, \emph{no 
renormalization} of the strength of the \emph{real} folded potential (\ref{eq11}) 
was allowed in the present OM analysis to test its proximity to the real nucleon OP 
as well as the impact of the RT to the real folded OP. While the imaginary folded 
nucleon OP based on the JLM parametrization of the G-matrix interaction delivers 
a good OM description of elastic proton scattering, it gives consistently a stronger
absorption in the neutron OP, and an overall renormalization of the \emph{imaginary} 
neutron folded potential by a factor $\sim 0.8$ is needed for a good OM description 
of elastic neutron scattering data at the considered energies. From the OM results 
obtained with the CDM3Y6 interaction shown in Fig.~\ref{f3} for elastic \nPb scattering 
one can see that the inclusion of the RT into the folding model calculation is essential 
for a good OM description of the data over the whole angular range. We note that 
the results obtained with the complex CDM3Y3 and CDM3Y4 interactions are almost 
the same as those shown in Fig.~\ref{f3}, with a minor difference 
that is hardly noticeable on the logarithmic scale. 
\begin{figure}[bht] \vspace*{0cm}\hspace*{2cm}
\includegraphics[width=0.8\textwidth]{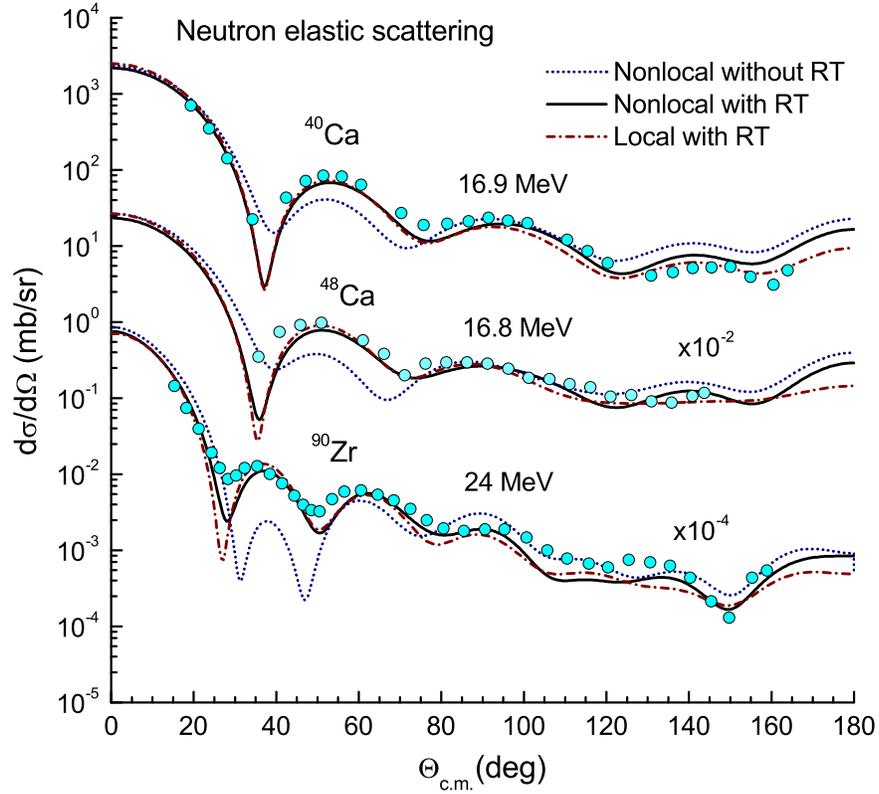}\vspace*{-0.5cm}
 \caption{The same as Fig.~\ref{f3} but for the data of the elastic neutron 
scattering on $^{40,48}$Ca and $^{90}$Zr target measured \cite{ndat1,ndat2,ndat3} 
at 17 and 24 MeV, respectively.} \label{f4}
\end{figure}
In the absence of the Coulomb interaction, the oscillation pattern of the elastic
neutron cross section over the whole angular range can be reproduced only with the 
inclusion of the RT. It can also be seen in Fig.~\ref{f3} that the local approximation 
(\ref{eq20})-(\ref{eq21}) for the exchange term of the folded nucleon OP is quite 
reasonable, and the local folded OP with the RT included also gives a reasonable OM 
description of the data. About the same impact by the RT and good accuracy of the 
local folding approach can be seen in the OM results for elastic neutron scattering 
on the medium-mass $^{40,48}$Ca and $^{90}$Zr targets (see Fig.~\ref{f4}).   

\begin{figure}[bht] \vspace*{-1cm}\hspace*{2cm}
\includegraphics[width=0.8\textwidth]{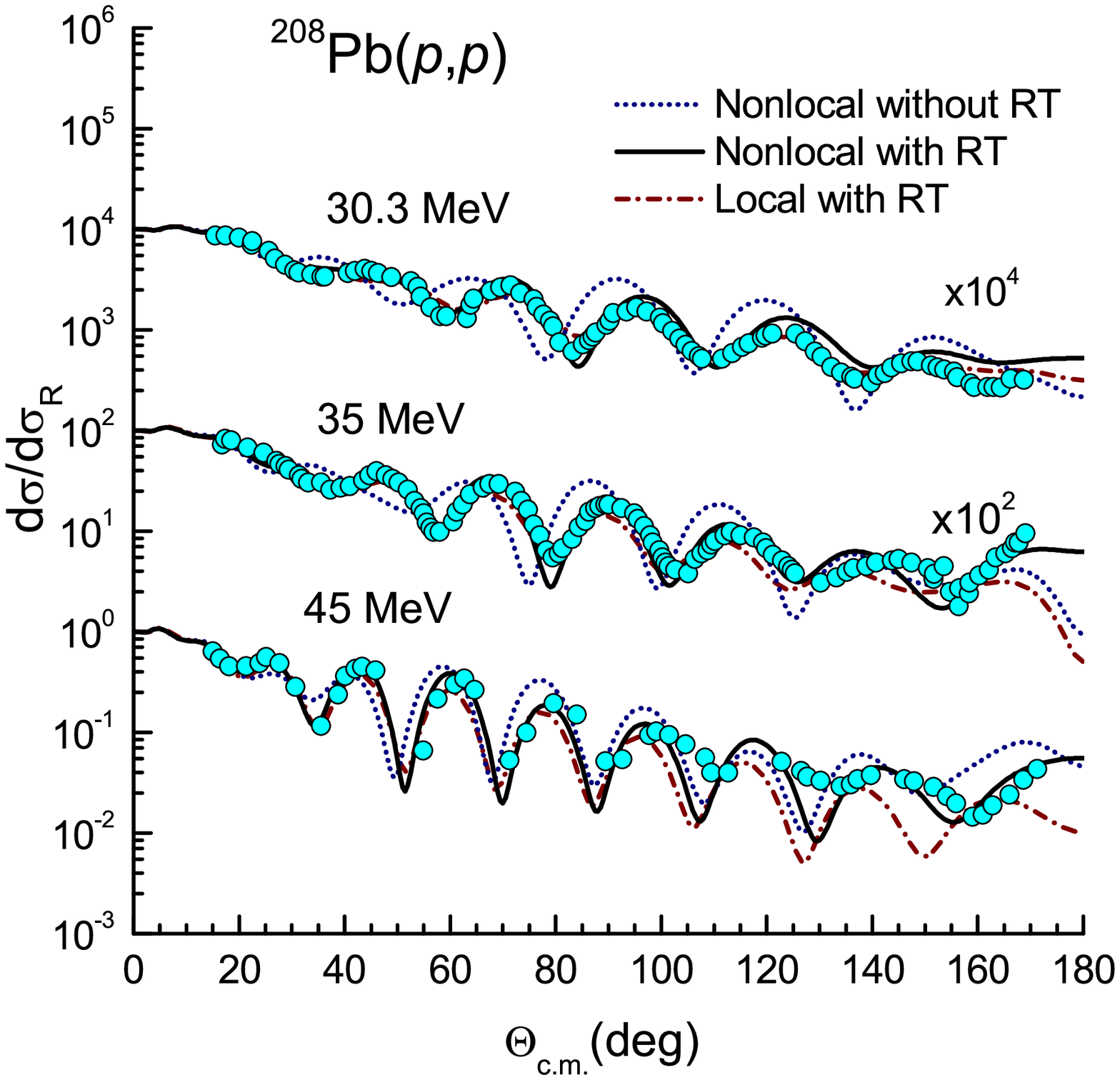}\vspace*{-1cm}
 \caption{The same as Fig.~\ref{f3} but for the elastic \pPb scattering data 
 measured at 30, 35, and 45 MeV \cite{pPb1,pPb2}. The calculated elastic 
scattering cross sections and data points are plotted in ratio to the 
Rurtherford cross section at the corresponding angles.} \label{f5}
\end{figure}
\begin{figure}[bht] \vspace*{-1cm}\hspace*{2cm}
\includegraphics[width=0.8\textwidth]{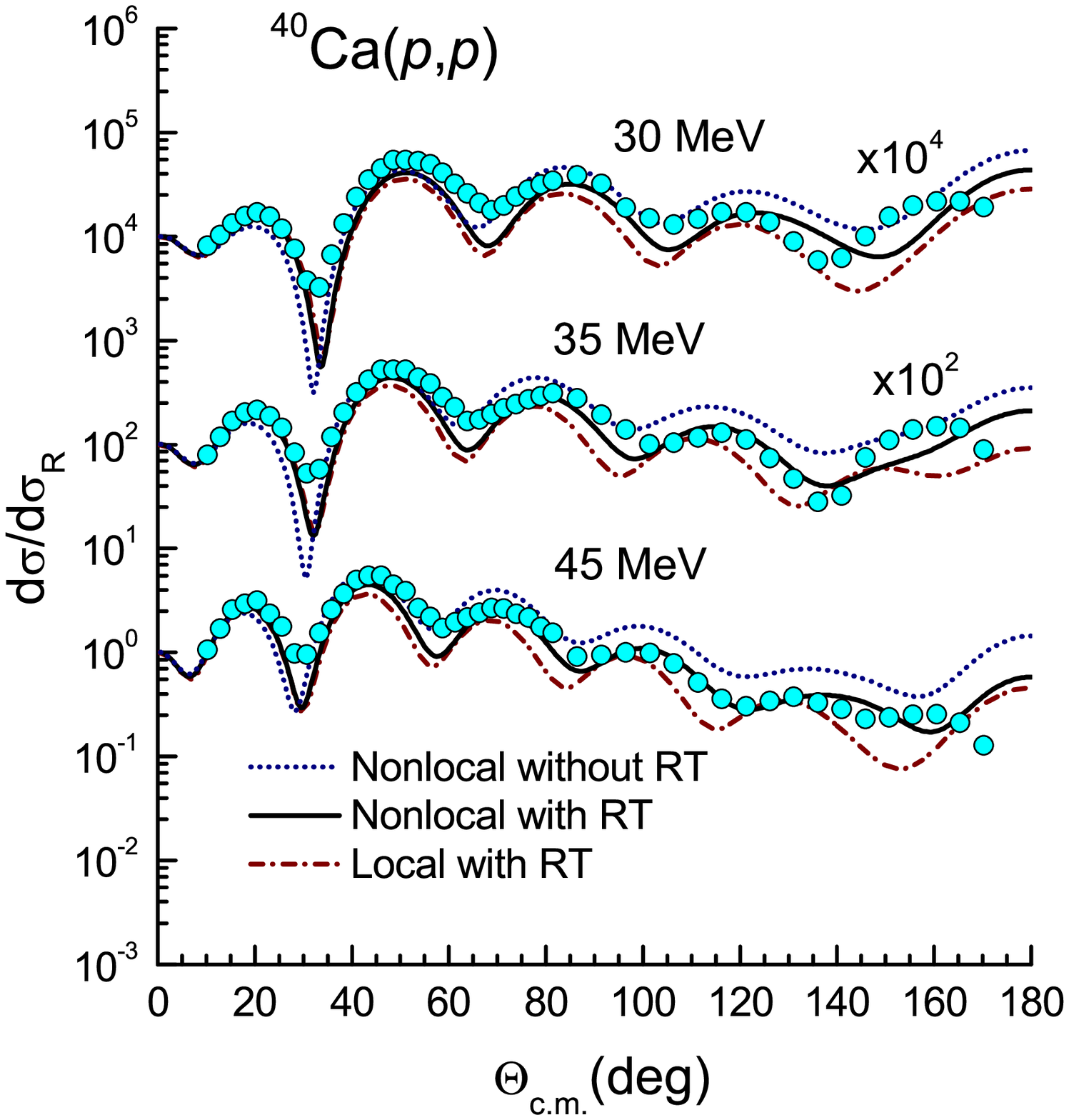}\vspace*{-1cm}
 \caption{The same as Fig.~\ref{f5} but for the elastic \pCa scattering data 
 measured at 30, 35, and 45 MeV \cite{nCa}.} \label{f6}
\end{figure}
The elastic \pPb scattering data measured at 30.4, 35, and 45 MeV \cite{pPb1,pPb2} 
are compared in Fig.~\ref{f5} with the OM results given by the same three versions 
of the folded OP as those discussed in Fig.~\ref{f3}. The inclusion of the RT into
the folding model calculation was found to be also vital for a good OM description 
of the elastic proton scattering data over the whole angular range as is seen  
in Fig.~\ref{f5}. At the forward angles, the effect of the RT in the elastic \pPb 
scattering is slightly weaker than that found in the elastic \nPb scattering because 
the elastic cross section there has a significant contribution from the Coulomb 
scattering which is not affected by the inclusion of the RT in the folding 
calculation. The same impact by the RT, but with a more pronounced difference 
between the results given by three versions of the folded OP can be 
seen in the OM results for the elastic \pCa scattering shown in Fig.~\ref{f6}.
For this double magic target, the results of our folding model analysis show 
unambiguously the importance of taking into account both the RT and nonlocality 
of the folded nucleon OP. Although, the nonlocal and local folded OP's give 
about the same OM results for the elastic proton scattering at the forward- 
and medium angles, the data at the backward angles can be properly reproduced 
only by the nonlocal folded OP, especially at the proton energy of 45 MeV. 
The effect of the RT found here in the folding model description of elastic \nA 
scattering at low energies should be complementary to the rearrangement of the SP 
configurations established in the single-nucleon removal reactions \cite{Hodgson75}.  

\section{Microscopic nonlocal folded OP versus the global parametrization}
\label{sec5} 
\begin{figure}[bht] \vspace*{-1cm}
\includegraphics[width=1.1\textwidth]{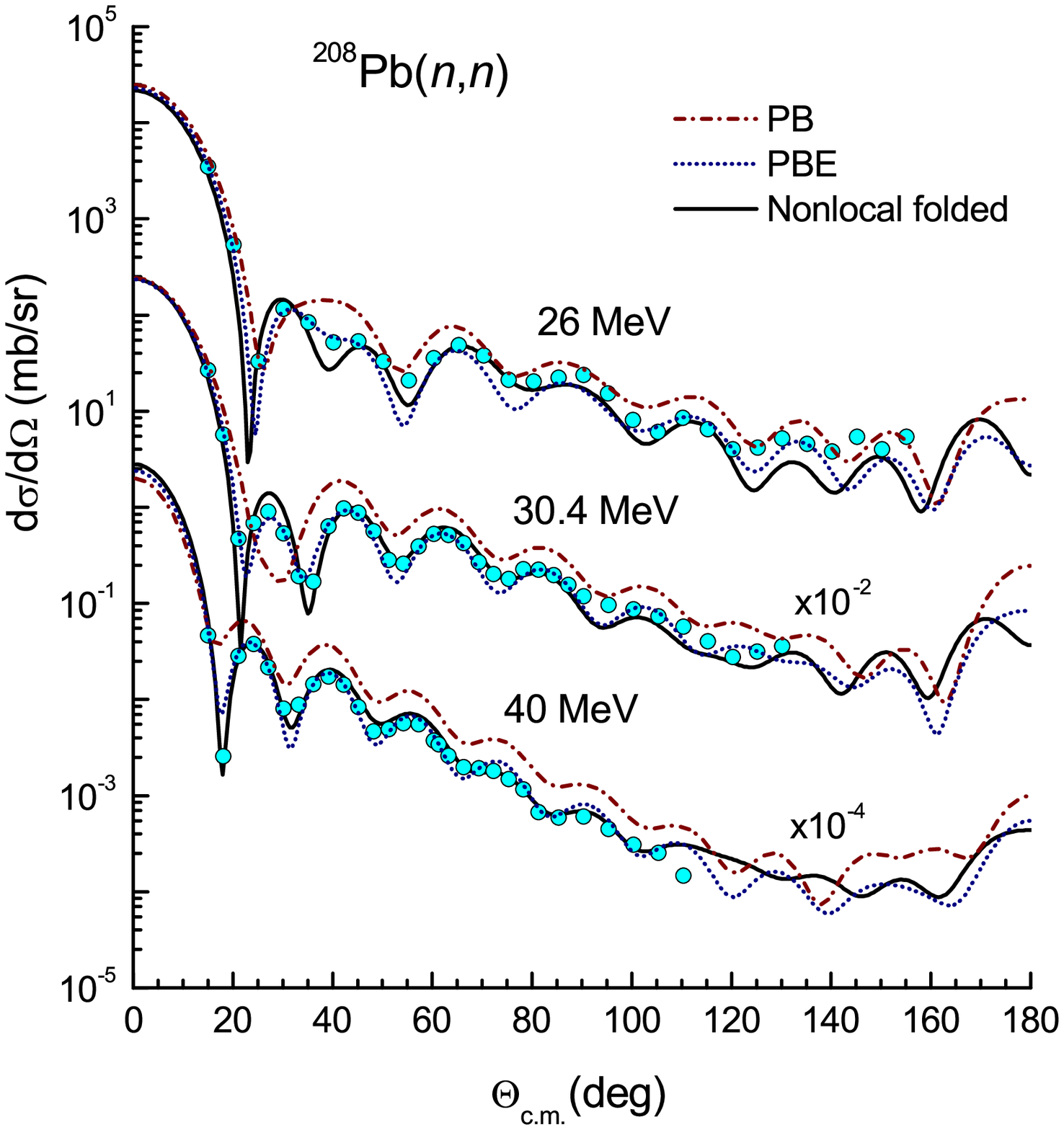}\vspace*{-3.5cm}
 \caption{OM description of the elastic \nPb scattering data measured at 26, 30, 
and 40 MeV \cite{nPb1,nPb2,nPb3} given by the nonlocal folded OP obtained with 
the CDM3Y6 interaction including the RT, in comparison with the OM results 
given by the original PB parametrization of the nonlocal neutron OP \cite{Perey} 
and its recent energy dependent version PBE \cite{Lovel}.} \label{f7}
\end{figure}
Although the practical OM calculations of elastic nucleon scattering are usually
done using some global parameters of the local nucleon OP (see, e.g., 
Refs.~\cite{Var91,Kon03}), some OM studies have been aimed to explore the use 
of an explicit nonlocal nucleon OP and deduce the global parameters for that 
purpose. We note here the early work by Perey and Buck (PB) \cite{Perey} and 
the recent revision of the PB parametrization by Tian, Pang, and Ma (TPM) \cite{TPM}, 
where the nonlocal nucleon OP is built up from a Woods-Saxon form factor multiplied 
by a nonlocal Gaussian. While the PB parameters were adjusted to the best OM fit 
of the two data sets (elastic \nPb scattering at 7.0 and 14.5 MeV), those 
of the TPM potential were fitted to reproduce the data of elastic nucleon scattering 
on $^{32}$S, $^{56}$Fe, $^{120}$Sn, and $^{208}$Pb targets at energies of 8 to 30 MeV. 
More recently, an energy dependence has been introduced explicitly into the imaginary 
parts of the PB and TPM potentials, dubbed as PBE and TPME potentials, with the
parameters adjusted to achieve the overall good OM description of nucleon elastic 
scattering on $^{40}$Ca, $^{90}$Zr, and $^{208}$Pb targets at energies $E$ of 
5 to 45 MeV \cite{Lovel,Lovel2}.
\begin{figure}[bht] \vspace*{-1cm}
\includegraphics[width=1.1\textwidth]{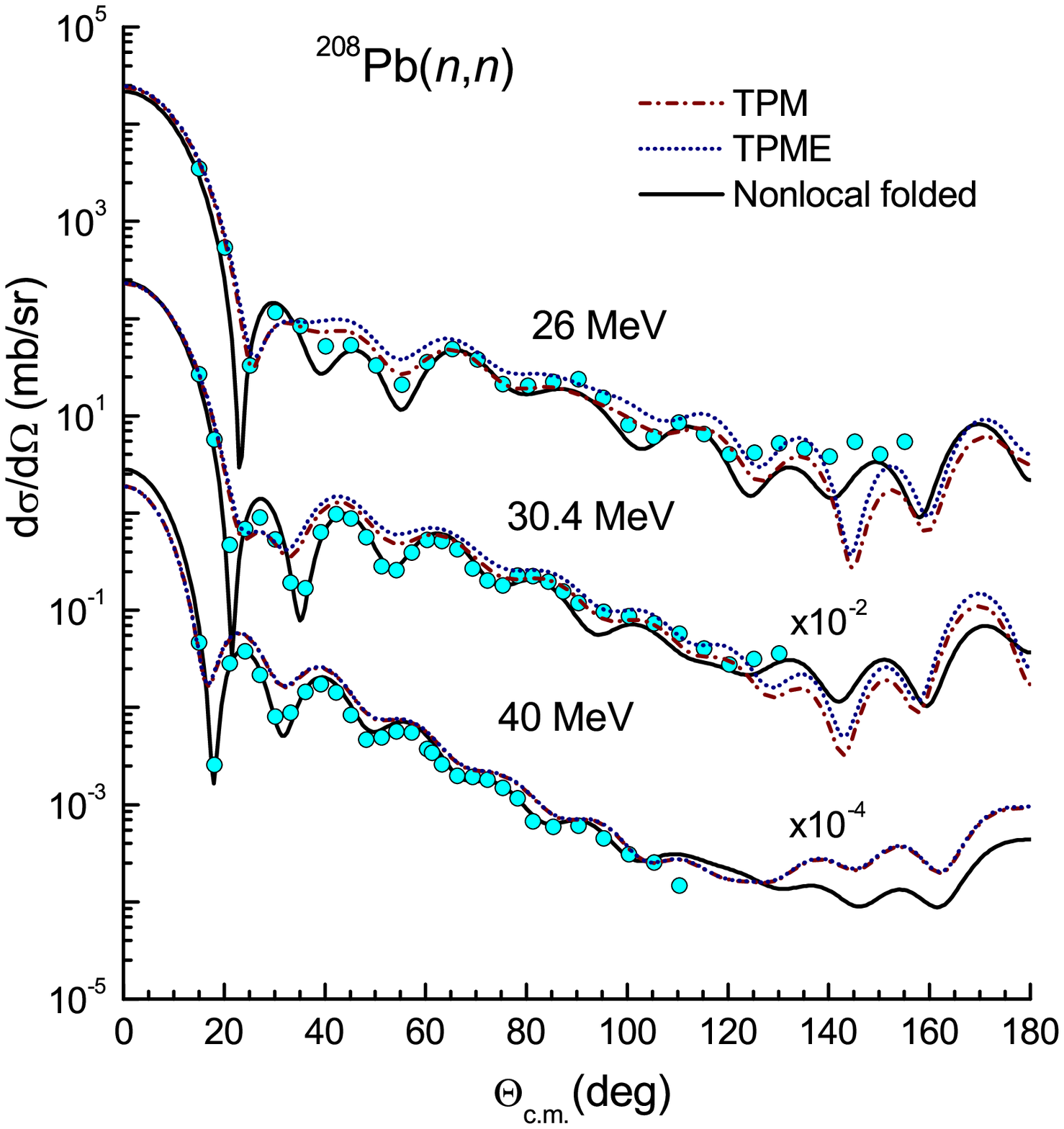}\vspace*{-3.5cm}
 \caption{The same as Fig.~\ref{f7} but in comparison with the OM results 
 given by the TPM parametrization of the nonlocal neutron OP \cite{TPM} 
 and its energy dependent version TPME \cite{Lovel}.} \label{f8}
\end{figure}
Given the microscopic nucleon folded OP constructed from the realistic SP wave 
functions of the target nucleons using the mean-field based density dependent CDM3Yn
interaction, it is of interest to compare its predicting power with that of the 
global parametrization. The OM results for the elastic \nPb scattering at 26, 30.4, 
and 40 MeV given by the nonlocal folded OP obtained with the CDM3Y6 interaction are 
compared with the OM results given by the PB parametrization of the nonlocal 
neutron OP \cite{Perey} and the recent energy dependent version PBE \cite{Lovel} 
in Fig.~\ref{f7}. One can see that the nonlocal folded OP performs quite well, 
with the predicted elastic cross section agreeing closely with the data like 
that given by the global PBE potential. 
As expected, the original (energy independent) PB parametrization fails to 
account for these data that were measured at energies higher than those considered
by Perey and Buck \cite{Perey}. It can be seen in Fig.~\ref{f8} that the elastic 
\nPb cross section predicted by the nonlocal folded OP agrees with the data 
slightly better than that predicted by the global TPM and TPME parametrizations 
of the nonlocal OP. Thus, these results confirm nicely the reliability of the 
present nonlocal version of the folding model in the calculation of the nucleon OP 
for medium and heavy targets. The model can be used, therefore, to predict 
the complex nucleon OP of the short-lived, unstable nuclei (when no elastic 
scattering data are available) for further use in the direct reaction analysis,
provided the realistic SP wave functions are available for these radioactive nuclei,
which are not an easy task for the nuclear structure models.   

\section{Summary}
The folding model of the nonlocal nucleon OP, with the exchange potential 
calculated exactly in the HF manner, is generalized to include the rearrangement
term using the CDM3Yn interaction with a complex density dependence. The obtained 
OM results for the elastic neutron and proton scattering on $^{40,48}$Ca, $^{90}$Zr, 
and $^{208}$Pb targets at different energies have shown that the inclusion of the RT 
into the folding model calculation of the nucleon OP is essential for a good OM 
description of the considered elastic data. Although the RT is widely taken into 
account in numerous variational HF calculations of nuclear structure using the effective
density-dependent NN interaction, the results of the present work have confirmed, for 
the first time, the important role of the RT in the (local and nonlocal) HF-type 
folding model of the nucleon OP.  

The OM results given by the complex nonlocal folded OP with the RT included are 
further compared with those given by the global parametrization of the nonlocal 
nucleon OP suggested recently \cite{TPM,Lovel}, based on the analytical nonlocal form 
factor suggested many years ago by Perey and Buck \cite{Perey}. This comparison 
has confirmed the reliability of the present nonlocal folding model for the medium 
and heavy targets, and the model can be used, therefore, to predict the nucleon OP 
of the short-lived unstable nuclei (for which no elastic scattering data are available) 
using the realistic SP wave functions given by the nuclear structure studies. 

\section*{Acknowledgments}
We thank Pierre Descouvemont for his helpful discussions and comments on the 
calculable $R$-matrix method \cite{desco,desco2}. The present research 
has been supported, in part, by the National Foundation for Scientific 
and Technological Development (NAFOSTED Project No. 103.04-2017.317).

\end{document}